\documentclass[aps,prd,preprintnumbers]{revtex4-2}
\pdfoutput=1
\usepackage{graphicx}
\usepackage{amsmath,amssymb}
\usepackage[colorlinks]{hyperref}

\begin{document}
\title{Local correlation among the chiral condensate, monopoles, and color magnetic fields
in Abelian projected QCD}

\author{Hideo~Suganuma}
\email{suganuma@scphys.kyoto-u.ac.jp} 
\affiliation{Department of Physics, Kyoto University,
Kitashirakawaoiwake, Sakyo, Kyoto 606-8502, Japan}

\author{Hiroki~Ohata}
\email{hiroki.ohata@yukawa.kyoto-u.ac.jp} 
\affiliation{Yukawa Institute for Theoretical Physics,
Kyoto University, Sakyo, Kyoto 606-8502, Japan}

\date{\today}
\begin{abstract}
Using the lattice gauge field theory, we study 
the relation among the local chiral condensate, monopoles, 
and color magnetic fields in quantum chromodynamics (QCD).
First, we investigate idealized Abelian gauge systems of 
1) a static monopole-antimonopole pair and 2) a magnetic flux without monopoles, 
on a four-dimensional Euclidean lattice.
In these systems, we calculate the local chiral condensate on quasi-massless fermions 
coupled to the Abelian gauge field, and find that the chiral condensate is localized 
in the vicinity of the magnetic field. 
Second, using SU(3) lattice QCD Monte Carlo calculations, 
we investigate Abelian projected QCD in the maximally Abelian gauge, 
and find clear correlation of distribution similarity
among the local chiral condensate, monopoles, and color magnetic fields in the Abelianized gauge configuration.
As a statistical indicator, 
we measure the correlation coefficient $r$, 
and find a strong positive correlation of $r \simeq 0.8$ 
between the local chiral condensate and an Euclidean 
color-magnetic quantity ${\cal F}$ in Abelian projected QCD.
The correlation is also investigated 
for the deconfined phase in thermal QCD.
As an interesting conjecture, like magnetic catalysis, the chiral condensate is locally enhanced by the strong color-magnetic field around the monopoles in QCD. 
\end{abstract}

%\keyword{QCD; chiral symmetry; monopole; lattice QCD; spontaneous symmetry breaking; Abelian projection; magnetic catalysis} 

\preprint{YITP-21-85}
\maketitle

\section{Introduction}

Quantum chromodynamics (QCD) is an SU($N_c$) gauge theory 
to describe the strong interaction, 
and has presented many interesting subjects full of variety and difficult problems in physics.  
Actually, in spite of the simple form of the QCD action, 
this miracle theory creates hundreds of hadrons and 
leads to various interesting nonperturbative phenomena, 
such as color confinement 
and dynamical chiral-symmetry breaking~\cite{Nambu:1961tp}.

This magic is due to the strong coupling of QCD 
in the low-energy region, and this strong-coupling nature 
drastically changes the vacuum structure itself.
Therefore, perturbative technique is no more workable and 
analytical treatment of QCD is fairly difficult in the strong-coupling region. 
As a reliable standard technique, 
lattice QCD Monte Carlo simulations have been applied to analyze nonperturbative QCD~\cite{Creutz:1980zw,Rothe:2012nt}. 

Among the nonperturbative properties of QCD, 
spontaneous chiral-symmetry breaking is particularly 
important in our real world. 
Indeed, chiral symmetry breaking drastically 
influences the vacuum structure 
and gives a nontrivial vacuum expectation value of  
the chiral condensate $\langle \bar{q}q \rangle$, 
which plays the role of an order parameter.
Also, it is considered that chiral symmetry breaking 
leads to dynamical quark-mass generation~\cite{Nambu:1961tp,Higashijima:1983gx},
and creates most of the matter mass of our Universe, 
apart from the dark matter, because 
only small masses of u, d current quarks and electrons 
are Higgs-origin in atoms~\cite{Zyla:2020zbs} 
and their contribution to the nucleon mass 
is estimated to be small~\cite{Ji:1995sv}.
In addition, chiral symmetry breaking inevitably accompanies 
light pions of the Nambu-Goldstone bosons, 
and their small mass gives range of the nuclear force.

In nonperturbative QCD, 
color confinement is also one of the most important phenomena 
in physics, 
and presents an extremely difficult mathematical problem.
Experiments for hadron spectra and lattice QCD studies 
for various inter-quark 
potentials~\cite{Takahashi:2000te,Takahashi:2002bw,Okiharu:2004ve,Okiharu:2004wy}
show that the quark confining force 
is basically characterized by a universal physical quantity 
of the string tension $\sigma \simeq$ 0.89 GeV/fm. 
This universal string tension is physically explained by 
one-dimensional squeezing of the color electric flux, 
i.e., the color flux-tube formation in hadrons, 
as is also indicated by lattice QCD for both mesons~\cite{Rothe:2012nt} and baryons~\cite{Bornyakov:2004uv}. 
As for the relation between color confinement and chiral symmetry breaking, it is not yet clarified directly from QCD. 
While almost coincidence between deconfinement and chiral-restoration
temperatures~\cite{Karsch:2001cy} 
suggests their close correlation, 
a lattice QCD analysis using the Dirac-mode expansion 
based on the Banks-Casher relation~\cite{Banks:1979yr}
indicates some independence of these phenomena in QCD~\cite{Doi:2014zea,Suganuma:2014wya}.

For the quark confinement mechanism, 
Nambu~\cite{Nambu:1974zg}, 't~Hooft~\cite{tHooft:1975krp}, and Mandelstam~\cite{Mandelstam:1974pi} proposed the dual superconductivity scenario, 
paying attention to analogy with the Abrikosov vortex in the superconductivity, 
where Cooper-pair condensation leads to the Meissner effect, 
and the magnetic flux is excluded or squeezed like a 
one-dimensional tube as the Abrikosov vortex.
%
%where the magnetic flux is quantized topologically. 
%
If the QCD vacuum can be regarded as the dual version of the superconductor, 
the electric-type color flux is squeezed between (anti)quarks in hadrons,
and quark confinement can be physically explained by the dual Meissner effect.
%
%the low-dimensionalization of the quantized flux between electric-type charges . 
Because of the electromagnetic duality, 
the dual Meissner effect inevitably needs 
condensation of magnetic objects, i.e., color magnetic monopoles, 
which correspond to the dual version of the electric Cooper-pair bosonic field.

In the dual-superconductor picture for the QCD vacuum, 
however, there are two large gaps with QCD.
\begin{enumerate}
\item
While QCD is a non-Abelian gauge theory, 
the dual-superconductor picture is based on an Abelian gauge theory subject to the Maxwell-type equations including magnetic currents, where electromagnetic duality is manifest.
\item
While QCD includes only color electric variables, i.e., quarks and gluons, as the elementary degrees of freedom, 
the dual-superconductor picture requires condensation of 
color magnetic monopoles as a key concept.
\end{enumerate}
Historically, to bridge between QCD and the dual-superconductivity, 
't~Hooft proposed Abelian gauge fixing~\cite{tHooft:1981bkw},
partial gauge fixing which only remains 
Abelian gauge degrees of freedom in QCD.
By Abelian gauge fixing, QCD reduces into an Abelian gauge theory, 
where off-diagonal gluons behave as charged matter fields 
similar to $W^\pm_\mu$ in the Weinberg-Salam model 
and give the color electric current $j_\mu$ 
in terms of the residual Abelian gauge symmetry.
As a remarkable fact in the Abelian gauge, 
color-magnetic monopoles appear as topological objects 
corresponding to the nontrivial homotopy group 
$\Pi_2({\rm SU}(N_c)/{\rm U}(1)^{N_c-1})={\bf Z}^{N_c-1}_\infty$ 
in a similar manner to appearance of 
't~Hooft-Polyakov monopoles \cite{Shnir:2011zz}
in the SU(2) non-Abelian Higgs theory.
%In general, the monopole appears as a topological defect or 
%a singularity in a constrained Abelian gauge manifold embedded 
%in the compact (and at most semi-simple) non-Abelian gauge manifold.
%
Thus, in the Abelian gauge, 
QCD is reduced into an Abelian gauge theory  
including both electric current $j_\mu$ and 
magnetic current $k_\mu$, 
which is expected to give a theoretical basis 
of the dual-superconductor picture for the confinement mechanism,
although off-diagonal gluons remain as charged matter fields.

From the viewpoint of Abelianzation of QCD,
the maximally Abelian (MA) gauge~\cite{Kronfeld:1987ri} 
is an interesting special Abelian gauge.
In the MA gauge, off-diagonal gluons have 
a large effective mass of about 1 GeV in 
both SU(2) and SU(3) lattice QCD~\cite{Amemiya:1998jz,Gongyo:2012jb,Gongyo:2013sha}, 
so that off-diagonal gluons become infrared inactive,
and only the Abelian gluon is relevant 
at distances larger than about 0.2 fm. 
%
%Then, in the MA gauge, 
%infrared QCD becomes an Abelian gauge theory 
%including the magnetic monopole current $k_\mu$  
%together with the electric current $j_\mu$. 
%
Also, monopole condensation is suggested from 
appearance of 
long entangled monopole worldlines ~\cite{Kronfeld:1987ri,Kronfeld:1987vd} 
and the magnetic screening in lattice QCD~\cite{Suganuma:2000jh,Hoelbling:2000su}.

In this way, by taking the MA gauge, 
the QCD vacuum can be regarded as an Abelian dual superconductor 
at a large scale, 
and color magnetic monopoles seem to 
capture essence of nonperturbative QCD.
Note however that, even without gauge fixing, 
there is an evidence of monopole condensation
in nonabelian gauge theories~\cite{Hoelbling:2000su}, 
and therefore it might be possible 
to define infrared-relevant monopoles in QCD and 
to construct the dual superconductor system 
in more general manner. 
%characterized by monopole condensation.
%
In fact, MA gauge fixing gives a concrete way 
to extract infrared-relevant Abelian gauge manifold 
and monopoles from QCD.

In the context of the dual superconductir picture, 
close correlation between monopoles and chiral symmetry breaking 
was pointed out in the dual Ginzburg-Landau theory~\cite{Suganuma:1993ps},
in SU(2) lattice QCD in the MA gauge~\cite{Miyamura:1995xn,Woloshyn:1994rv}, 
and in SU(3) lattice QCD~\cite{Lee:1995ac,Thurner:1997qx}.
Since most of the pioneering lattice studies were done 
in SU(2) QCD or on a small lattice as $8^3\times 4$, 
we recently investigated SU(3) QCD with a large volume, 
and find a clear correlation between monopoles and the chiral condensate in SU(3) lattice QCD in the MA gauge~\cite{Ohata:2020myj}.

In this paper, 
as a continuation of Ref.\cite{Ohata:2020myj},
we proceed the lattice works 
for the relation between chiral symmetry breaking
and color magnetic objects including monopoles.
In particular, as a new point of this paper, 
we quantitatively study correlation of the local chiral condensate with color magnetic fields using 
the lattice gauge theory.
%
%we investigate not only monopoles but also color 
%magnetic fields as a new research.

The organization of this paper is as follows. 
In Section 2, we review the MA gauge and 
Abelianization of QCD in SU(3) lattice formalism.
In Section 3, we prepare 
magnetic objects in Abelian projected QCD.
In Section 4, we consider the local chiral condensate and chiral symmetry breaking in Abelian gauge systems.
In Section 5, we present idealized 
Abelian gauge systems of a static monopole-antimonopole pair on a lattice, 
and investigate the relation of the local chiral condensate with the magnetic objects.
In Section 6, we perform SU(3) lattice QCD Monte Carlo calculations 
and study the relation among monopoles, magnetic fields, 
and the local chiral condensate in Abelian projected QCD in the MA gauge.
Section 7 is devoted for summary and conclusion.

\section{Maximally Abelian gauge and Abelianization of QCD}

To begin with, we briefly review  
the lattice formalism for 
maximally Abelian (MA) gauge fixing and
Abelianization in QCD. 

Continuum QCD is described with 
the quark field $q(x)$, the gluon field $A_{\mu}(x) \in {\rm su}(N_c)$ and 
the QCD gauge coupling $g$.
In SU($N_c$) lattice QCD~\cite{Rothe:2012nt}, 
the gluon field is described as 
the SU($N_c$) link variable 
$U_{\mu}(s) \equiv \exp\left( i a g A_{\mu}(s) \right) \in {\rm SU}(N_c)$
on four-dimensional Euclidean lattices 
with the spacing $a$ and the 
%space-time 
volume $V =L_x L_y L_z L_t$.

%Here, we briefly explain MA gauge fixing 
%and Abelian projection in SU(3) lattice QCD.
Using the Cartan subalgebra $\vec{H} \equiv (T_3,T_8)$ in SU(3), MA gauge fixing is defined so as to maximize 
\begin{equation}
  R_{\rm MA}[U_{\mu}(s)] \equiv \sum_s \sum_{\mu=1}^4 {\rm tr} 
  \left( U_{\mu}^\dag(s) \vec{H} U_{\mu}(s) \vec{H} \right) 
  = \sum_{s} \sum_{\mu=1}^4 
  \left( 1 - \frac{1}{2}\sum_{i \neq j} 
\left| U_{\mu}(s)_{ij} \right|^2 \right) \label{eq:Abelian}
\end{equation}
by the SU(3) gauge transformation, 
and therefore this gauge fixing strongly suppresses 
all the off-diagonal fluctuation of the SU(3) gauge field. 
In the MA gauge, the SU(3) gauge group is partially fixed remaining 
its maximal torus subgroup ${\rm U(1)}_3 \times {\rm U(1)}_8$ 
with the global Weyl (color permutation) symmetry~\cite{Ichie:1999hb},
and QCD is reduced to an Abelian gauge theory.
%like the non-Abelian Higgs theory.

From the SU(3) link variable $U_{\mu}^{\rm MA}(s) \in {\rm SU(3)}$ in the MA gauge, we extract  
the Abelian link variable 
%\begin{equation}
%  u_{\mu}(s) = \exp \left\{ i \theta_{\mu}^3(s) T_3 + i \theta_{\mu}^8(s) T_8 \right\} \in 
%{\rm U}(1)_3 \times {\rm U}(1)_8
%\end{equation}
\begin{equation}
  u_{\mu}(s) = e^{i\vec \theta_\mu(s) \cdot \vec H}
  =\mathrm{diag}\left( e^{i\theta_{\mu}^{1}(s)}, e^{i\theta_{\mu}^{2}(s)},
  e^{i\theta_{\mu}^{3}(s)} \right) 
  \in {\rm U}(1)_3 \times {\rm U}(1)_8 \subset {\rm SU}(3)
  \label{eq:Abelian_field} 
\end{equation}
by maximizing the overlap
\begin{equation}
  R_{\rm Abel} \equiv 
  \frac{1}{3} {\rm Re \, tr} \left\{ U_{\mu}^{\rm MA}(s) u_{\mu}^\dag(s) \right\}
  \in \left[ -\frac{1}{2}, 1 \right].
\end{equation}
Note that the distance between $u_{\mu}(s)$ and $U_{\mu}^{\rm MA}(s)$
becomes the smallest in the SU(3) manifold, 
and there is a constraint $\sum_{i=1}^{3} \theta_{\mu}^{i}(s) = 0\, \left( \mathrm{mod}~2\pi \right)$
reflecting the uni-determinant of $u_{\mu}(s)$.
Here, $\theta_\mu^i (s)$ ($i$ = 1, 2, 3) is taken to be the principal value of $-\pi \le \theta_\mu^i (s)< \pi$.

The Abelian projection is defined 
by the simple replacement of the SU(3) link variable $U_\mu(s)$ 
by the Abelian link variable $u_\mu(s)$ for each gauge configuration, that is, 
$O[U_\mu(s)] \rightarrow O[u_\mu(s)]$ for QCD operators. 
Abelian projected QCD is thus extracted from SU(3) QCD.
The case of $\langle O[U_\mu(s)] \rangle \simeq \langle O[u_\mu(s)] \rangle$ 
is called ``Abelian dominance'' for the operator $O$ 
\cite{Ezawa:1982bf}. 

As a remarkable fact, 
Abelian dominance of quark confinement is 
shown in both SU(2)~\cite{Suzuki:1989gp}
and SU(3) lattice QCD~\cite{Sakumichi:2014xpa,Sakumichi:2015rfa,Ohata:2020dgn}.
Also, Abelian dominance of chiral symmetry breaking is
observed in SU(2)~\cite{Miyamura:1995xn,Woloshyn:1994rv,Lee:1995ac}
and SU(3) lattice QCD~\cite{Ohata:2020myj}.

\section{Magnetic Objects in Abelian projected QCD}
\label{Sect:magnetic}

In this section, we prepare 
magnetic objects in Abelian projected QCD
in four-dimensional Euclidean space-time.

\subsection{Monopoles in Abelian projected QCD}

In this subsection, we define monopoles in 
Abelian projected QCD in lattice formalism~\cite{DeGrand:1980eq}.
Like the ordinary SU(3) plaquette, 
the Abelian plaquette variable is defined as 
\begin{eqnarray}
u_{\mu\nu}(s) 
&\equiv&
u_{\mu}(s) u_{\nu}(s+\hat \mu) u_{\mu}^\dagger(s+\hat \nu)  u_{\nu}^\dagger(s) 
=e^{i\vec \theta_{\mu\nu}(s) \cdot \vec H} \cr
&=&{\rm diag}(e^{i\theta_{\mu\nu}^{1}(s)},e^{i\theta_{\mu\nu}^{2}(s)}, e^{i\theta_{\mu\nu}^{3}(s)}) \in {\rm U(1)}^2
\subset {\rm SU}(3),
\end{eqnarray}
where $\hat \mu$ is the $\mu$-directed unit vector in the lattice unit.
The Abelian field strength $\theta_{\mu \nu}^{i}(s)$ ($i=1,2,3$) is 
the principal value of the exponent in $u_{\mu\nu}(s)$, and is defined as 
\begin{align}
  \partial_{\mu} \theta_{\nu}^{i}(s) - \partial_{\nu}\theta_{\mu}^{i}(s) 
  &= \theta_{\mu\nu}^{i}(s) - 2\pi n_{\mu\nu}^{i}(s), \notag \\
  -\pi \le \theta_{\mu\nu}^{i}(s) &< \pi, \quad n_{\mu\nu}^{i}(s) \in \mathbb{Z},
\end{align}
with the forward derivative $\partial_{\mu}$. 
Note that $\theta_{\mu\nu}^{i}(s)$ is ${\rm U(1)}^2$ gauge invariant 
and corresponds to the regular Abelian field strength in the continuum limit of $a \rightarrow 0$, 
while $n_{\mu\nu}^{i}(s)$ corresponds to 
the singular gauge-variant Dirac string~\cite{DeGrand:1980eq}.

The electric current $j_\mu^i$ and the monopole current $k_\mu^i$ are 
defined from the Abelian field strength $\theta_{\mu\nu}^i$ as 
\begin{eqnarray}
j_\nu^i(s) &\equiv& \partial_\mu^{\prime} \theta_{\mu\nu}^i(s), 
\\
k_\nu^i(s) &\equiv& \partial_\mu \tilde{\theta}_{\mu\nu}^i(s)/2\pi
=\partial_\mu \tilde{n}_{\mu\nu}^i(s) \in \mathbb{Z},
\end{eqnarray}
where $\partial_{\mu}^{\prime}$ is the backward derivative 
and $\tilde{\theta}_{\mu\nu}$ is the dual tensor of 
$\tilde{\theta}_{\mu\nu} \equiv \frac12 \epsilon_{\mu\nu\alpha\beta}{\theta}_{\alpha\beta}$.
Both electric and monopole currents are $\mathrm{U}(1)^2$ gauge invariant, 
according to $\mathrm{U}(1)^2$ gauge invariance of ${\theta}_{\mu\nu}^i(s)$.
In the lattice formalism, 
$k_{\mu}^{i}(s)$ is located at the dual lattice $L_{\rm dual}^4$
of $s^\alpha+\frac{1}{2}$ with flowing in $\mu$ direction~\cite{Ichie:1998ns}.
%Hereafter, we will omit the color index $i$ as %appropriate. 

In this way, Abelian projected QCD includes both 
electric current $j_\mu^i$ and monopole current $k_\mu^i$. 
Remarkably, lattice QCD shows monopole dominance, i.e., 
dominant role of monopoles for quark confinement 
in the MA gauge~\cite{Stack:1994wm}. 
Also, lattice QCD shows monopole dominance for 
chiral symmetry breaking, that is, 
monopoles in the MA gauge crucially contribute to spontaneous chiral-symmetry breaking
in both SU(2)~\cite{Miyamura:1995xn,Lee:1995ac} 
and SU(3) lattice QCD~\cite{Ohata:2020myj}. 

In the lattice formalism, the monopole current $k_\mu^i$ 
appears on the dual lattice $L_{\rm dual}^4$ of $s^\alpha+\frac{1}{2}$, 
and therefore we define the local monopole density 
\begin{eqnarray}
\rho_{\mathrm{L}}(s) \equiv \frac{1}{3 \cdot 2^4} 
\sum_{i=1}^{3}
\sum_{s'\in P(s)}
\sum_{\mu=1}^4
\left| k_{\mu}^{i}(s')
 \right|,
\end{eqnarray}
where $P(s)$ denotes the dual lattices in the vicinity of $s$, i.e.,
$P(s)=\left\{s'\in L_{\rm dual}^4 \middle| 
|s-s'|=1 \right\}$.
Note here that the distance between the site $s$ and 
its closest dual site $s'$ is 
$|s-s'|=\sqrt{\sum_1^4 (\frac{1}{2})^2}=1$ 
in the four-dimensional Euclidean space-time.
%
%Considering the possibility of cancellation of 
%monopole charges on the dual lattice, 
%we define also the local monopole charge density
%\begin{eqnarray}
%\rho_{\mathrm{LC}}(s) \equiv \frac{1}{3 \cdot 2^4} 
%\sum_{i=1}^{3}
%\sum_{s'\in P(s)}
%\sum_{\mu=1}^4 k_{\mu}^{i}(s') \right.
%\end{eqnarray}

\subsection{General argument for magnetic instability 
and magnetic objects in the QCD vacuum}

In QCD in the MA gauge, color magnetic monopoles generally appear, 
and play an important role in nonperturbative properties,
which migth looks curious, since the original QCD action does not have monopoles. 

However, some active roles of magnetic objects would be natural in QCD,
because QCD itself has color magnetic instability,
and spontaneous generation of color magnetic fields generally takes place, 
as Savvidy first pointed out in 1977~\cite{Savvidy:1977as,Nielsen:1978rm}. 

In fact, in the QCD vacuum in the Minkowski space-time, 
the gluon condensate
$\langle G_{\mu\nu}^aG_{\mu\nu}^a \rangle$
takes a large positive value, which physically 
means that the QCD vacuum is filled with color magnetic fields.
Since the gluon condensate is expressed 
with color magnetic fields $\vec H_a$ 
and color electric fields $\vec E_a$ as  
\begin{eqnarray}
\langle G_{\mu\nu}^aG_{\mu\nu}^a \rangle 
=2 (\langle \vec H_a^2\rangle -
\langle \vec E_a^2 \rangle) > 0, 
\end{eqnarray} 
its large positivity means inevitable significant generation of color magnetic fields.
Thus, some superior role of magnetic objects is expected 
instead of electric objects in the Minkowski QCD vacuum.

In the Euclidean space-time, 
because of the space-time $SO(4)$ symmetry, 
the roles of magnetic and electric fields become similar.
Actually, the gluon condensate is written as
$G_{\mu\nu}^aG_{\mu\nu}^a  
=2 (\vec H_a^2+\vec E_a^2)$, 
where the electromagnetic duality is manifest.  
Then, in Euclidean QCD, the electric field often behaves as 
a magnetic field, 
and therefore we regard the Euclidean electric field as a sort of the magnetic field in this paper. 

\subsection{Lorentz invariant quantities in Abelian projected QCD}

In Abelian projected QCD, there are two 
Lorentz invariant quantities ${\cal F}$ and ${\cal G}$
in the Euclidean space-time: 
\begin{align}
{\cal F} &\equiv \frac{1}{3} \sum_{i=1}^3 \frac{1}{4} F^{\mu\nu}_i F^{\mu\nu}_i 
= \frac{1}{3} \sum_{i=1}^3 \frac{1}{2} (\vec H_i^2+ \vec E_i^2), \\
{\cal G} &\equiv \frac{1}{3} \sum_{i=1}^3 \frac{1}{4} F^{\mu\nu}_i \tilde F^{\mu\nu}_i 
= \frac{1}{3} \sum_{i=1}^3 \ \vec H_i \cdot \vec E_i,
\end{align}
with the color magnetic field $(\vec H_i)_j \equiv \frac12 \epsilon_{jkl}F_i^{kl}$ 
and the color electric field 
$(\vec E_i)_j \equiv F_i^{j4}$.
These quantities are also invariant under 
the residual U(1)$^2$ gauge transformation 
and global Weyl transformation~\cite{Ichie:1999hb}, i.e., 
permutation of the color index, in the MA gauge.

Here, ${\cal F}$ is parity-even and 
expresses total magnitude 
of magnetic fields in the Euclidean space-time,
since the electric field behaves as a magnetic field there.
In this paper, we simply call ${\cal F}$ 
``magnetic quantity'' in Euclidean gauge theories. 
Note that ${\cal G}$ is parity-odd and 
is just the Abelian projected quantity 
of the topological charge density on instantons in QCD,
which might relate to chiral symmetry breaking.

In the lattice formalism, 
the field strength tensor is 
a plaquette variable spanning at 
$s$, $s+\hat \mu$, $s+\hat \nu$, and $s+\hat \mu+\hat \nu$,  
so that we define the Abelian field strength
$F_{\mu\nu}^i(s) $
as the local average of clover-type four plaquettes,
\begin{eqnarray}
a^2gF_{\mu\nu}^i(s) 
\equiv \frac14 
\left( \theta^i_{\mu\nu}(s)+\theta^i_{\mu\nu}(s+\hat \mu)+\theta^i_{\mu\nu}(s+\hat \nu)+
\theta^i_{\mu\nu}(s+\hat \mu +\hat \nu) \right),
\end{eqnarray}
and consider ${\cal F}$ and ${\cal G}$ 
as local quantities 
in each Abelian gauge configuration.

\section{Local chiral condensate and chiral symmetry breaking in gauge theories}

In this section, we consider 
the chiral condensate and chiral symmetry breaking
in the gauge theory in terms of the quark propagator.

\subsection{Local chiral condensate in lattice QCD}

In this subsection, we briefly review  
the local chiral condensate 
in lattice formalism. 
The local chiral condensate can be calculated 
with the quark propagator
for each gauge configuration 
$U = \{U_\mu(s)\}$ generated with the Monte Carlo method.

As the lattice fermion, we here adopt 
the Kogut-Susskind (KS) fermion~\cite{Rothe:2012nt}.
For the KS fermion, 
the Dirac operator $\gamma_\mu D_\mu$ is expressed by 
$\eta_\mu D_\mu$ with the staggered phase 
$\eta_\mu(s)\equiv (-1)^{s_1+\cdots+s_{\mu-1}}$ 
$(\mu \geq 2)$ 
with $\eta_1(s)\equiv 1$.
The KS Dirac operator is expressed as 
\begin{eqnarray}
(\eta_\mu D_\mu)_{ss'}
%=\frac{1}{2}\sum_{\mu=1}^{4}\eta_\mu(s)
%[U_\mu(s)\delta_{s+\hat{\mu},s'}
%-U_{-\mu}(s) \delta_{s-\hat{\mu},s'}]\cr
=\frac{1}{2}\sum_{\mu=1}^{4}\sum_{\pm}
\pm \eta_\mu(s)
U_{\pm \mu}(s)\delta_{s \pm \hat{\mu},s'}
\end{eqnarray}
with $U_{-\mu}(s) \equiv U^\dagger_\mu(s-\hat \mu)$,
and 
the KS Dirac eigenvalue equation takes the form of
%\begin{eqnarray}
%\frac{1}{2}\sum_{\mu=1}^4 &&
%\eta_\mu(s)[U_\mu(s) \chi_n(s+\hat \mu)-U_{-\mu}(s)
%\chi_n(s-\hat \mu)] \cr
%&=&i\lambda_n\chi_n(s). 
%\end{eqnarray}
\begin{eqnarray}
\frac{1}{2}\sum_{\mu=1}^4 \sum_{\pm} 
\pm \eta_\mu(s) U_{\pm\mu}(s) \chi_n(s\pm\hat \mu)
=i\lambda_n\chi_n(s). 
\end{eqnarray}
%In the KS fermion formalism, 
Here, the quark field $q^\alpha(s)$ is described by 
a spinless Grassmann variable $\chi(s)$ \cite{Rothe:2012nt}, 
and the chiral condensate per flavor is evaluated as  
$\langle \bar qq\rangle=\langle \bar \chi \chi\rangle$/4 in the continuum limit.

The local chiral condensate can be calculated 
using the quark propagator of  
the KS fermion with a small quark mass $m$. 
The chiral-limit value is estimated 
by the chiral extrapolation of $m \rightarrow 0$. 
As a technical caution, 
the chiral and continuum limits do not commute 
for the KS fermion at the quenched level, 
although this problem would be absent in full QCD 
\cite{Bernard:2004ab}.

For the gauge configuration $U=\{U_\mu(s)\}$, 
the Euclidean KS fermion propagator is given by
\begin{eqnarray}
G_U^{ij}(x,y) \equiv \langle {\chi}^i(x) \bar \chi^j(y) \rangle_U 
=\langle x,i|\left(
\frac{1}{\eta_\mu D_\mu[U]+m}\right)|y,j\rangle
\end{eqnarray}
with the color index $i$ and $j$.
This propagator is numerically obtained 
by solving the large-scale linear equation 
with a point source. 
The local chiral condensate 
for the gauge configuration $\{U_\mu(s)\}$
is expressed with the propagator as 
\begin{eqnarray}
\langle \bar \chi(x)\chi(x) \rangle_U=-{\rm Tr}\, G_U(x,x).
\end{eqnarray}
Here, we consider the net chiral condensate by subtracting  
the contribution from the trivial vacuum $U = 1$ as
\begin{eqnarray}
\langle \bar \chi\chi(x) \rangle_U
\equiv 
\langle \bar \chi(x)\chi (x)\rangle_U
-\langle \bar \chi\chi \rangle_{U=1},
\end{eqnarray}
where the subtraction term is exactly zero 
in the chiral limit $m=0$. 
The global chiral condensate 
is obtained by taking its average
over the space-time $x$
and the gauge ensembles $U_1, U_2, ..., U_N$,
\begin{eqnarray}
\langle \bar \chi\chi\rangle
\equiv \sum_{x,i}
\langle \bar \chi\chi (x)\rangle_{U_i}/\sum_{x,i} 1.
\end{eqnarray}

\subsection{Chiral symmetry breaking in Abelian gauge system}
\label{subsec:analytic}

In this subsection, we analytically investigate  
relation between chiral symmetry breaking 
and the field strength 
in Euclidean Abelian gauge systems.
For the simple argument, 
we consider Euclidean U(1) gauge systems with 
quasi-massless Dirac fermions coupled to the U(1) gauge field, 
although it is straightforward
to generalize this argument to Abelian projected QCD 
with ${\rm U}(1)^2$ gauge symmetry.

In the U(1) gauge system, 
the chiral condensate is proportional to 
the functional trace of the fermion propagator,
\begin{eqnarray}
I \equiv {\rm Tr} \frac{1}{\not D+m}
=-m {\rm Tr}\frac{1}{D^2-m^2+\frac{g}{2}\sigma \cdot F},
\end{eqnarray}
with the covariant derivative 
$D_\mu \equiv \partial_\mu+igA_\mu$,
the field strength $F_{\mu\nu}$, 
$\sigma \cdot F \equiv \sigma_{\mu\nu} F_{\mu\nu}$ and
$\sigma_{\mu\nu} \equiv \frac{i}{2}[\gamma_\mu, \gamma_\nu]$.
Note that $D^2-m^2$ is a negative-definite operator, and all of its eigenvalues are negative.
Since the trace of any odd-number product of 
$\gamma$-matrices is zero, we find 
\begin{eqnarray}
I =-m {\rm Tr}\frac{D^2-m^2-\frac{g}{2}\sigma \cdot F}
{(D^2-m^2)^2-2g^2({\cal F}-\gamma_5 {\cal G})-\frac{g}{2}[D^2, \sigma \cdot F]},
\label{eq:csb}
\end{eqnarray}
with 
\begin{eqnarray}
{\cal F}\equiv \frac{1}{4} F_{\mu\nu}F_{\mu\nu}=\frac{1}{2} (\vec H^2+\vec E^2), \quad
{\cal G}\equiv \frac{1}{4} F_{\mu\nu}\tilde F_{\mu\nu}=
\vec H \cdot \vec E.
\end{eqnarray}
Because of the overall factor $m$ in $I$, 
$I$ goes to zero in the chiral limit $m\rightarrow 0$,
unless the denominator becomes zero in this limit.

Since the operator $(D^2-m^2)^2$ in the denominator 
is positive definite, to realize the zero denominator in $I$ in Eq.~(\ref{eq:csb}), 
we need a significant negative contribution from 
the other three terms including 
${\cal F}$, ${\cal G}$, or $[D^2, \sigma \cdot F]$.
For instance, in the absence of the field strength, i.e., $F_{\mu\nu}\equiv 0$, one finds 
near $m \simeq 0$ 
\begin{eqnarray}
I_{F_{\mu\nu} \equiv 0} 
=m {\rm Tr}\frac{1}{p^2+m^2}
= m \gamma \int \frac{d^4p}{(2\pi)^4}\frac{1}{p^2+m^2}
= m \frac{\gamma}{16\pi^2} \int^{\Lambda^2} dp^2 \frac{p^2}{p^2+m^2}
\simeq m \frac{\gamma\Lambda^2}{16 \pi^2}
\end{eqnarray}
with the UV cutoff $\Lambda$ 
and the degeneracy $\gamma$.
According to the positive denominator in the integrand,
$I_{F_{\mu\nu} \equiv 0}$ has no IR singularity 
to cancel $m$ of the numerator, 
and therefore $I_{F_{\mu\nu} \equiv 0}$ goes to zero in the chiral limit of $m\rightarrow 0$. 

To cancel $m$ in the numerator of $I$, we need 
a significantly large amount of the field strength 
so as to present zero mode 
in the denominator of $I$ 
and to keep $I$ non-zero in the chiral limit.
Note here that ${\cal F} (\ge 0)$ always gives 
a negative (non-positive) contribution in the denominator of $I$, while the contribution from 
${\cal G}$ or $[D^2, \sigma \cdot F]$ can be 
positive and negative.
In fact, the magnetic quantity ${\cal F}$ 
can give the zero mode in the denominator of $I$, 
even without the contribution from 
${\cal G}$ and $[D^2, \sigma \cdot F]$. 
In contrast, in Euclidean Abelian gauge systems, 
${\cal F}\equiv \frac{1}{4}F_{\mu\nu}^2=0$ 
means $F_{\mu\nu}=0$, and then 
${\cal G}=[D^2, \sigma \cdot F]=0$. 

To conclude, 
the magnetic quantity ${\cal F}$ 
is expected to be significantly important 
to realize chiral symmetry breaking 
in Euclidean Abelian gauge theories,
although, in some cases, the contribution from 
${\cal G}$ and $[D^2, \sigma \cdot F]$ 
can assist the realization of chiral symmetry breaking. 

In a special case of constant $F_{\mu\nu}$, one finds 
$[D^2, \sigma \cdot F]=0$ for the Abelian system,  
and obtains 
\begin{eqnarray}
I =-m {\rm Tr}\frac{(D^2-m^2)[(D^2-m^2)^2-2g^2{\cal F}]}
{[(D^2-m^2)^2-2g^2{\cal F}]^2-4g^2{\cal G}^2},
\end{eqnarray}
because of ${\rm tr} \gamma_5={\rm tr} \sigma_{\mu\nu}=
{\rm tr} \gamma_5 \sigma_{\mu\nu}=0$.
For more special case of a constant magnetic field,  
there occurs the Landau-level quantization, 
and the spatial degrees of freedom 
perpendicular to the magnetic field 
is frozen in the lowest Landau level. 
This infrared effective low-dimensionalization of 
the charged spinor dynamics 
induces chiral symmetry breaking in the chiral limit ~\cite{Suganuma:1990nn,Suganuma:1991ha,Klevansky:1992qe}, 
which is known as magnetic catalysis~\cite{Gusynin:1995nb}. 

\section{Abelian gauge system with a static monopole-antimonopole pair on a lattice}

In the QCD vacuum,
complicated monopole world-lines generally emerge in the MA gauge~\cite{Kronfeld:1987ri,Kronfeld:1987vd},
and therefore it is difficult to clarify 
the primary correlation with the chiral condensate
among the magnetic objects such as monopoles, ${\cal F}$ and ${\cal G}$.

In this section, 
to seek for the primary correlation with the chiral condensate,
we create idealized Abelian gauge system with 
a monopole-antimonopole pair on a lattice,  
and investigate the relation among the local chiral 
condensate, monoples, and magnetic fields.
Also, we consider a magnetic flux system without monopoles.  

For simplicity, we here consider 
U(1) lattice gauge systems 
described by U(1) link variables 
\begin{eqnarray}
u_\mu(s) = e^{i\theta_\mu(s)} \in {\rm U}(1),
\end{eqnarray}
and quasi-massless Dirac fermions coupled to U(1) gauge fields with the coupling $g=1$.

\subsection{Static monopole-antimonopole pair systems}

To begin with,
we deal with an idealized Abelian gauge system 
of a static monopole-antimonopole pair 
on a periodic lattice of the 
four-dimensional Euclidean space-time. 

In the three-dimensional space ${\bf R}^3$, 
let us consider a static monopole-antimonopole pair 
with the distance of $l$ in $z$-direction.
To realize such a lattice gauge system,  
we set the Abelian link-variable $u_\mu(s)$ to be 
\begin{eqnarray}
u_x(s)=u_y(s+\hat x)=u_x^\dagger(s+\hat y)=u_y^\dagger(s)=i
\quad  \hbox{for} \quad s_x=s_y=0, ~~ 1 \le s_z \le l,
\end{eqnarray}
otherwise $u_\mu(s)=1$.

\begin{figure}[htb]
\centering
\includegraphics[width = 10cm]{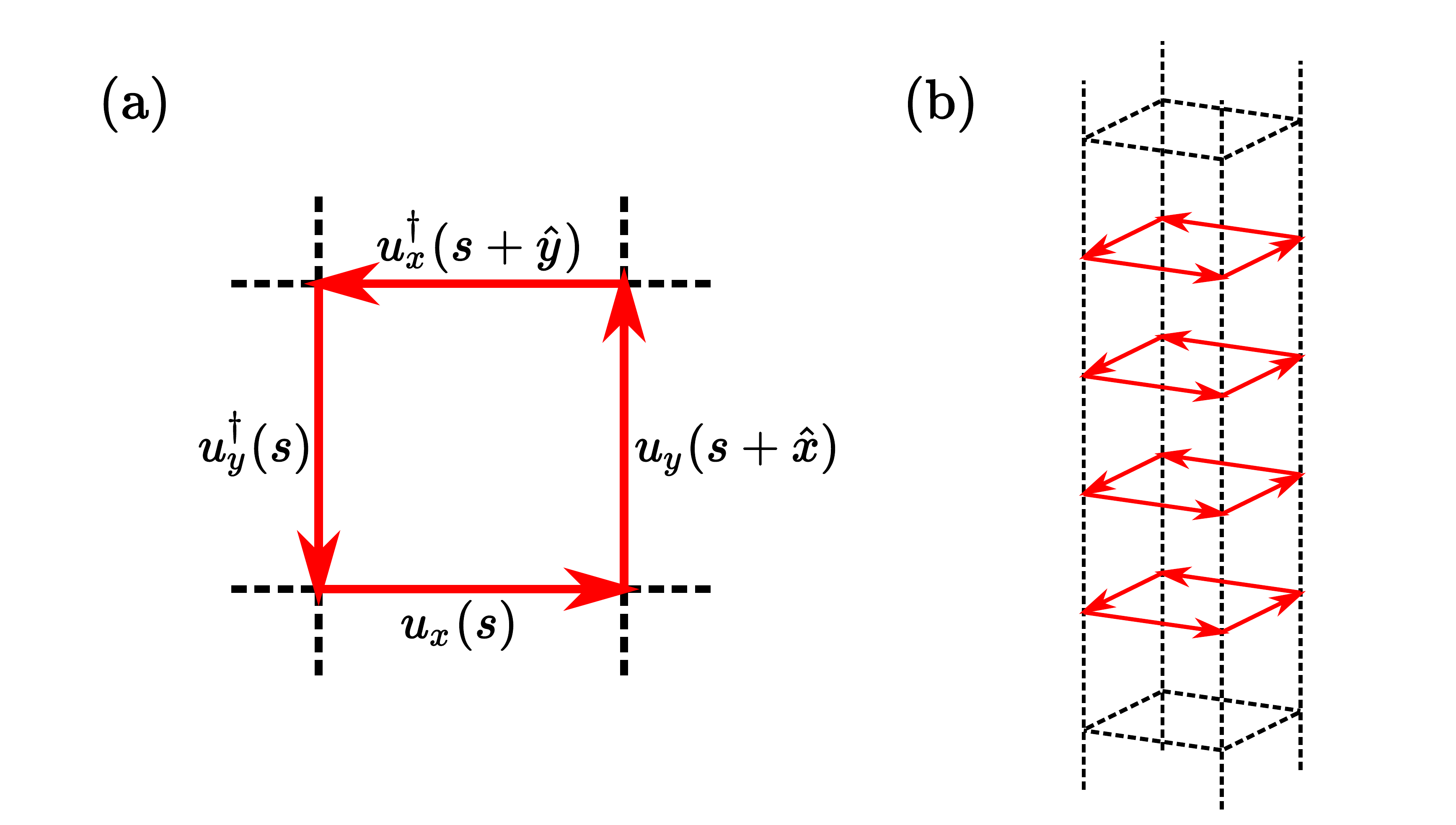}
\caption{The building-block plaquette 
to realize a static monopole-antimonopole pair on the lattice  
in (a) the $x$-$y$ plane and (b) spatial ${\bf R}^3$ 
for $s=(0, 0, s_z, s_t)$ with $1 \le s_z \le l$.
Only the red link-variables take a nontrivial value of $i$.
The all-red plaquette induces the singular Dirac string 
at its center on the dual lattice. 
A physical magnetic field is also created 
in the neighboring plaquette $u_{xy}(s)$ 
including only one red link.
}
\label{fig:schematic_artifi}
\end{figure}

Figure~\ref{fig:schematic_artifi} shows the building-block plaquette 
to realize a static monopole-antimonopole pair on the lattice. 
Here, only the red link-variables take a nontrivial value of $i$.

As for the phase variable $\theta_\mu(s)$, 
which corresponds to the Abelian gluon, 
one finds 
\begin{eqnarray}
\theta_x(s)=\theta_y(s+\hat x)=-\theta_x(s+\hat y)=-\theta_y(s)=\frac{\pi}{2}
\quad  \hbox{for} \quad s_x=s_y=0, ~~ 1 \le s_z \le l,
\end{eqnarray}
otherwise $\theta_\mu(s)=0$. 
For the all-red plaqutte with 
$s_x=s_y=0$ and $1 \le s_z \le l$,
one gets 
\begin{eqnarray}
  \partial_{x} \theta_{y}(s) - 
  \partial_{y}\theta_{x}(s) =
\theta_x(s)+\theta_y(s+\hat x)-\theta_x(s+\hat y)-\theta_y(s)=2\pi,
\end{eqnarray}
which leads to the Dirac string of $n_{xy}(s)=-1$ and 
zero field strength $\theta_{xy}(s)=0$,
%inside the all-red plaquette, 
because of the definition of the field strength $\theta_{\mu\nu}(s)$
and the Dirac string $n_{\mu\nu}(s)$, 
\begin{align}
  \partial_{\mu} \theta_{\nu}(s) - 
  \partial_{\nu}\theta_{\mu}(s) 
  = \theta_{\mu\nu}(s) - 2\pi n_{\mu\nu}(s), \quad
  -\pi \le \theta_{\mu\nu}(s) < \pi, \quad n_{\mu\nu}(s) \in \mathbb{Z}.
\end{align}
Thus, the all-red plaquette induces the singular Dirac string 
at its center on the dual lattice. 
In fact, for the idealized system in Fig.~\ref{fig:schematic_artifi}(b),
a Dirac string appears inside the all-red plaquette.

\begin{figure}[htb]
\centering
\includegraphics[width = 10 cm]{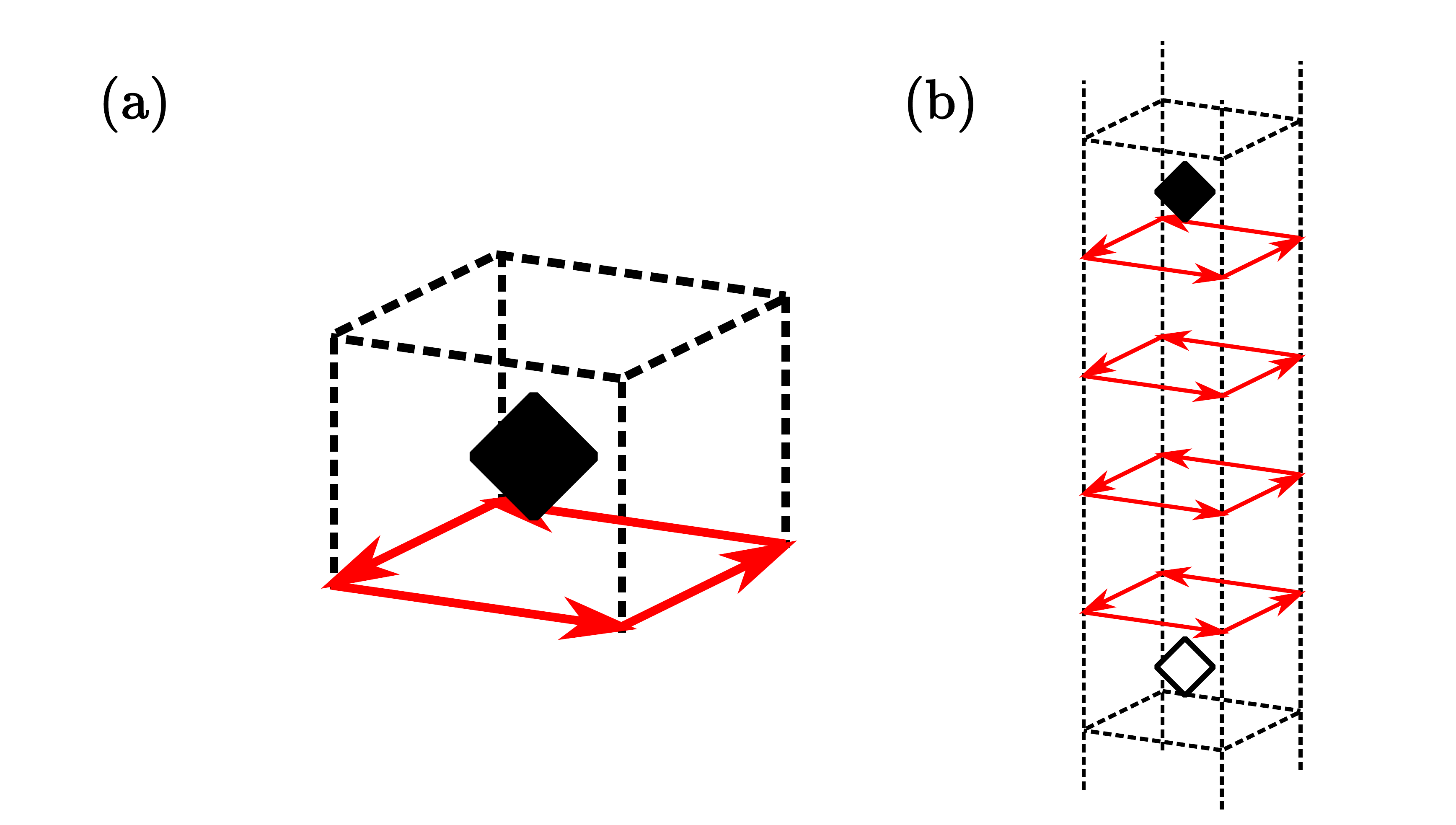}
\caption{The link-variables 
to realize a static monopole-antimonopole pair on the lattice 
in spatial ${\bf R}^3$.
Only the red link-variables take a nontrivial value of $i$.
(a) The cube including only one all-red plaquette 
induces a magnetic monopole at its inside on the dual lattice.
(b) A monopole (black diagmond) and an anti-monopole (white diamond)
appear at the two terminals of the red plaquette tower.
}
\label{fig:schematic_artifi2}
\end{figure}

At the terminal of the Dirac string, 
a monopole or an anti-monopole appears on the dual lattice, as shown in Fig.~\ref{fig:schematic_artifi2}.
Actually, the three-dimensional spatial cube including only 
one all-red plauette has a static (anti)monopole 
at its center (on the dual lattice), 
because only one $n_{kl}(s)$ has nonzero value of 
$\pm 1$ among the six independent plaquettes composing the cube,  
\begin{eqnarray}
k_4 (s) &=&\partial_j \tilde{n}_{j 4}(s)
=\frac12 \epsilon_{jkl} \partial_j n_{kl}(s)
=\frac12 \epsilon_{jkl} 
\{ n_{kl}(s+\hat j)-n_{kl}(s)\} \cr
&=& n_{xy}(s+\hat z)-n_{xy}(s)
+n_{yz}(s+\hat x)-n_{yz}(s)
+n_{zx}(s+\hat y)-n_{zx}(s) \cr
&=&\pm 1.
\end{eqnarray}
Thus, this idealized system includes a static monopole 
at $(\frac{1}{2}, \frac{1}{2}, \frac{1}{2})$ 
and a static anti-monopole at 
$(\frac{1}{2}, \frac{1}{2}, l+\frac{1}{2})$ 
in spatial ${\bf R}^3$. 

This monopole and anti-monopole system has 
also physical magnetic flux around 
the line segment connecting the monopole pair.
In fact, a physical magnetic field is created 
in the neighboring plaquette $u_{xy}(s)$ 
of the all-red plaquette 
in Fig.~\ref{fig:schematic_artifi}.
%
%with a Dirac string between them, and 
%a physical magnetic flux appears around 
%the Dirac string.
%
In this idealized system, 
only the plaquette $u_{xy}(s)$ 
including one red link takes a nontrivial value as
\begin{eqnarray}
u_{xy}(s)=-i=e^{-i\pi/2}
\quad  \hbox{in case with one nontrivial link},
\end{eqnarray}
otherwise $u_{\mu\nu}(s)=1$.
Note here that, by gauge transformation, 
the location of the Dirac string is generally changed,
but the physical field strength is never changed.

\begin{figure}[htb]
\centering
\includegraphics[width = 14cm]{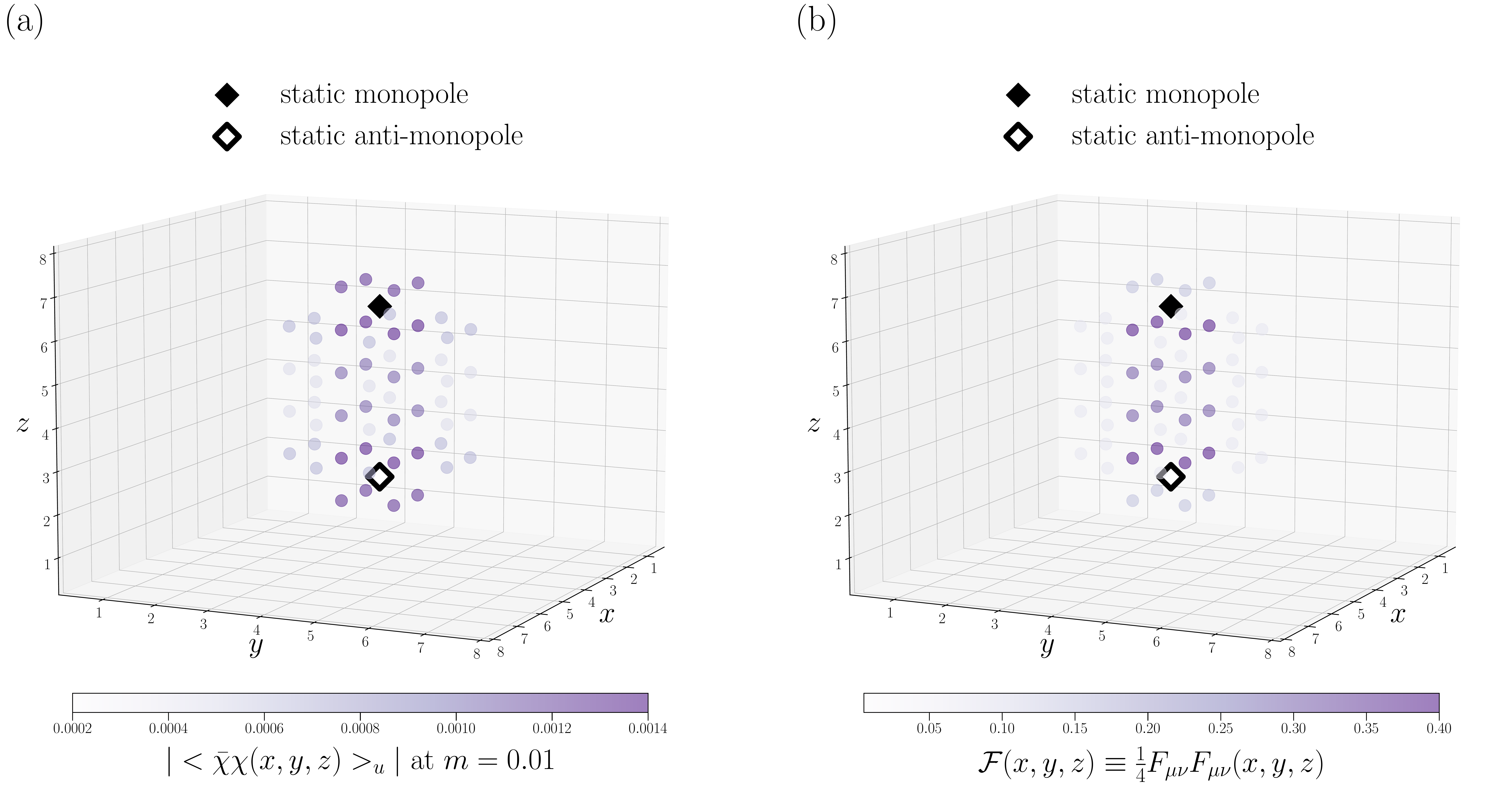}
\caption{
An idealized Abelian gauge system of  
a static monopole and anti-monopole pair 
with the distance of $l=4$. 
In the three dimensional space ${\bf R}^3$, 
the value is visualized with the color graduation for 
(a) the local chiral condensate and 
(b) the magnetic quantity ${\cal F}$. 
}
\label{fig:3dim_artifi}
\end{figure}

Figure~\ref{fig:3dim_artifi} shows 
the local chiral condensate 
$\langle \bar \chi \chi (s) \rangle_u$ 
and the magnetic quantity ${\cal F} \equiv \frac14 F_{\mu\nu}^2
=\frac12 \vec H^2$ for $l=4$ 
in the three dimensional space ${\bf R}^3$.
In this demonstration, the quark mass is taken to be 
$m=0.01$ in the lattice unit.

For this idealized static system, 
there actually appears 
a magnetic field $\vec H$, i.e.,  
nonzero flux of ${\cal F}=\frac{1}{2}\vec H^2>0$,
in space between the monopole and the anti-monopole, 
and the local chiral condensate 
takes a significant value 
in the vicinity of the magnetic field. 
In contrast, one finds 
${\cal G}= \vec H \cdot \vec E=0$ everywhere, 
since only spatial plaquettes take a nontrivial value 
and $\vec E = \vec 0$. 
Thus, in this system, it is likely that 
the magnetic field stemming from monopoles has the primary correlation with the local chiral condensate. 

\subsection{Static magnetic flux system}

Next, let us investigate a static magnetic flux system 
without monopoles. 
Owing to the spatial periodicity, the special case of $l=L_z$ 
in the static monopole-antimonopole system 
has no (anti)monopoles, because of the magnetic-charge cancellation. 
In this special case of $l=L_z$, 
there only exists a physical static magnetic flux 
along $z$-direction.

\begin{figure}[htb]
\centering
\includegraphics[width = 14cm]{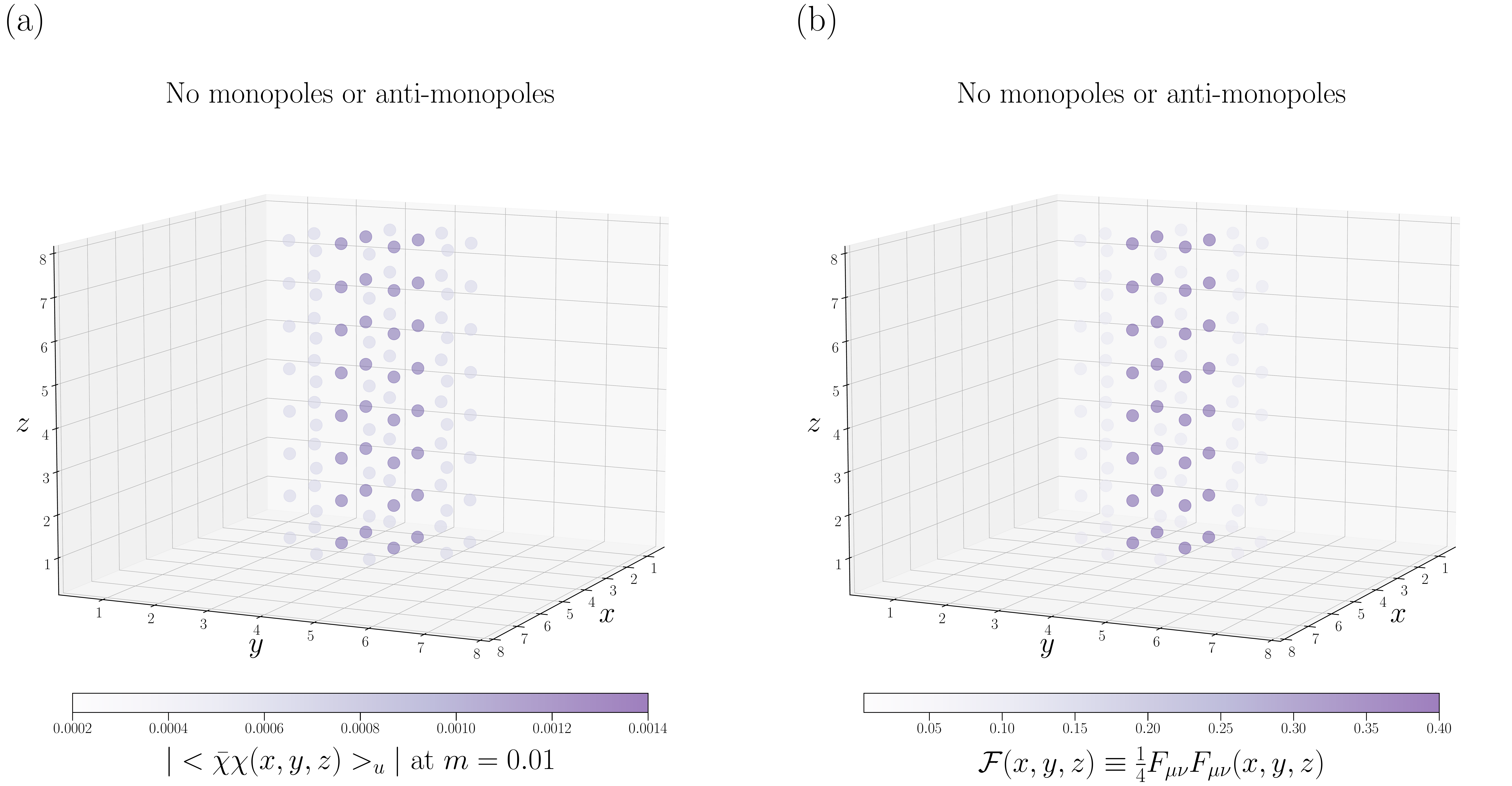}
\caption{
An idealized Abelian gauge system of  
a static magnetic flux without monopoles. 
In the three dimensional space ${\bf R}^3$, 
the value is visualized with the color graduation for 
(a) the local chiral condensate and 
(b) the magnetic quantity ${\cal F}$. 
}
\label{fig:3dim_artifi2}
\end{figure}

Figure~\ref{fig:3dim_artifi2} shows 
the local chiral condensate 
$\langle \bar \chi \chi (s) \rangle_u$ 
and the magnetic quantity ${\cal F} \equiv \frac14 F_{\mu\nu}^2
=\frac12 \vec H^2$ for $l=L_z$ 
in spatial ${\bf R}^3$, 
taking the quark mass of $m=0.01$ in the lattice unit.

Again, the local chiral condensate takes a significant value 
in the vicinity of the magnetic field $\vec H$, i.e.,  
nonzero flux of ${\cal F}=\frac{1}{2}\vec H^2>0$, 
even without (anti)monopoles. 
Note also that this system has ${\cal G}= \vec H \cdot \vec E=0$ everywhere, because of $\vec E = \vec 0$. 
Therefore, in this idealized system, we conclude that 
the magnetic field or ${\cal F}$ has the primary 
correlation with the local chiral condensate. 

\section{Lattice QCD study for local chiral condensate, monopoles, and magnetic fields}

In our previous study with lattice QCD, we observed 
a strong correlation between the local chiral condensate and monopoles in Abelian projected QCD~\cite{Ohata:2020myj}.
As a possible reason of this correlation, 
we conjectured that the strong magnetic field around monopoles is responsible to chiral symmetry breaking in QCD, similarly to the magnetic catalysis~\cite{Suganuma:1990nn,Suganuma:1991ha,Klevansky:1992qe,Gusynin:1995nb}.

In this section, 
using lattice QCD Monte Carlo simulations, 
we investigate the relation 
among the local chiral condensate, monopoles, 
and magnetic fields in Abelian projected QCD. 
In this paper, 
the SU(3) lattice QCD simulation is performed 
using the standard plaquette action 
at the quenched level.
In each space-time direction, we impose the periodic boundary 
condition for link variables, 
and the anti-periodic boundary condition for quarks 
in order to describe also thermal QCD.

For the numerical Monte Carlo calculation, 
we basically adopt the lattice parameter of 
$\beta \equiv 2N_c/g^2=6.0$ and the size $V = 24^4$. 
The lattice spacing $a \simeq$ 0.1 fm is 
obtained from the string tension $\sigma = 0.89$ GeV/fm 
\cite{Sakumichi:2014xpa}.
Also, we adopt $\beta=6.0$ and $V = 24^3 \times 6$ 
for the high-temperature deconfined phase 
at $T\simeq$ 330 MeV above the critical temperature.

Using the pseudo-heat-bath algorithm, 
we generate 100 and 200 gauge configurations 
for $V = 24^4$ and $24^3 \times 6$, respectively.
All the gauge configurations are taken every 500 sweeps 
after thermalization of 5,000 sweeps.
MA gauge fixing is performed with the stopping criterion 
that the deviation 
$\Delta R_{\mathrm{MA}} / \left( 4V \right)$ 
becomes smaller than $10^{-5}$ in 100 iterations.
For the calculation of the local chiral condensate, 
we use the quark propagator of the KS fermion 
with the quark mass of $m=0.01, 0.015, 0.02$ 
in the lattice unit,
%
%considering that 
Here, the quark mass is taken to be finite, since 
the chiral and continuum limits do not commute 
for the KS fermion at the quenched level 
\cite{Bernard:2004ab}.
The jackknife method is used for statistical error estimates.

For each lattice gauge configuration of 
Abelian projected QCD in the MA gauge, 
we calculate the local monopole density $\rho_{\rm L}(s)$, 
the local chiral condensate, and the Lorentz invariants 
${\cal F}$ and ${\cal G}$,  
defined in Section~\ref{Sect:magnetic}

\subsection{Distribution similarity between local chiral condensate and magnetic variables}

To begin with, we pick up a gauge configuration generated in 
lattice QCD on $V = 24^4$ at $\beta = 6.0$, 
and investigate correlation between 
the local chiral condensate and magnetic variables. 

\begin{figure}[htb]
\centering
\includegraphics[width=14 cm]{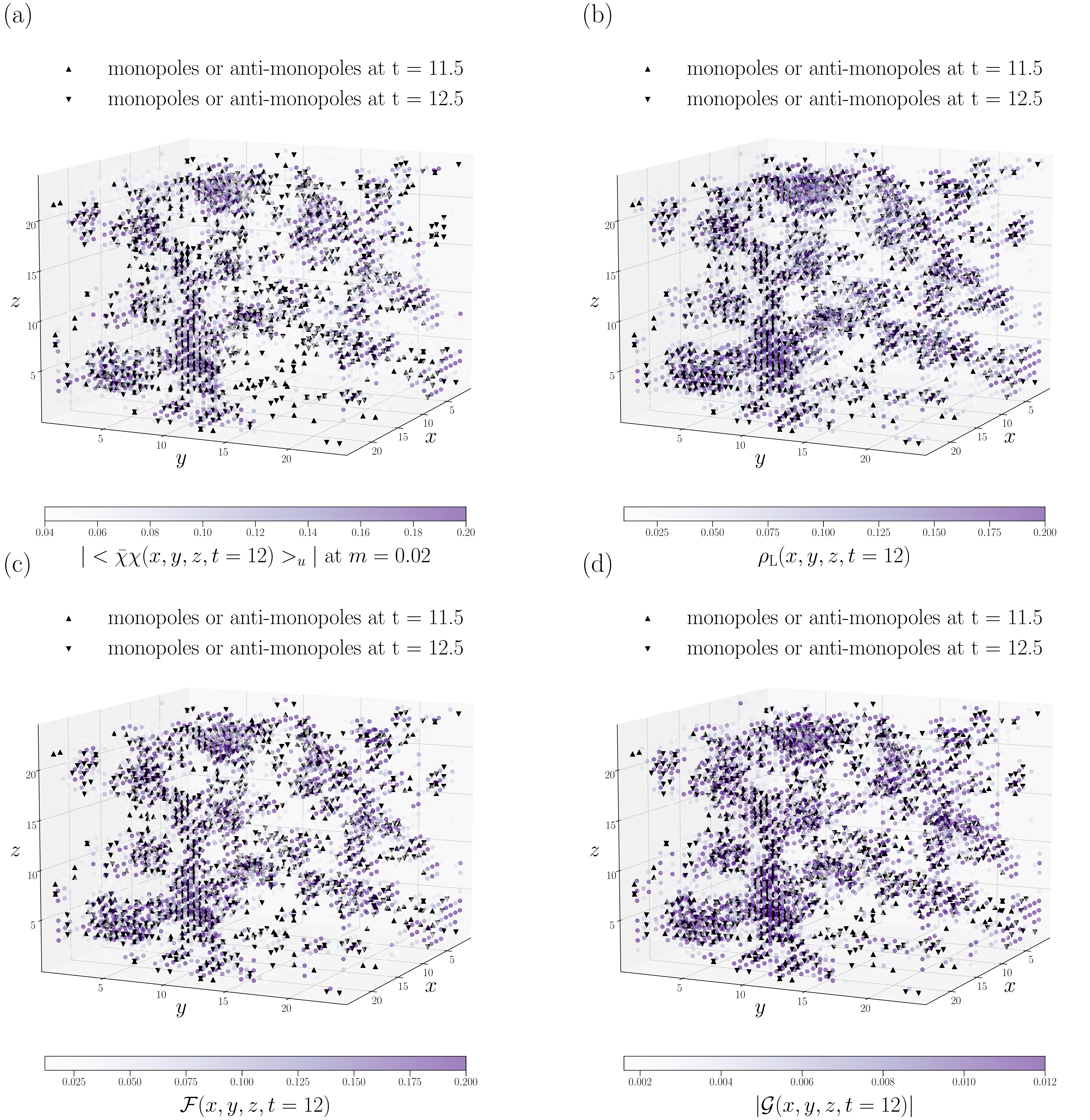}
\caption{Lattice QCD results for 
(a) the local chiral condensate $\langle \bar \chi \chi (s) \rangle_u$ with the quark mass of $m = 0.02$, 
(b) the local monopole density $\rho_{\rm L}(s)$, 
and the Lorentz invariants 
(c) ${\cal F}(s)$ and (d) $|{\cal G}(s)|$
in spatial ${\bf R}^3$ at a time slice, for 
a typical gauge configuration of Abelian projected QCD.
The value is visualized with the color graduation.
Monopoles at $t$ = 11.5 and 12.5 are plotted
with upper and lower triangles, respectively.
}
\label{fig:3dim}
\end{figure}

Figure~\ref{fig:3dim} shows the local chiral condensate $\langle \bar \chi \chi (s) \rangle_u$ with the quark mass of $m = 0.02$, 
the local monopole density $\rho_{\rm L}(s)$, 
and the Lorentz invariants 
${\cal F}(s)$ and $|{\cal G}(s)|$, respectively, 
as well as the monopole location in the space ${\bf R}^3$ 
at a time slice, for 
a typical gauge configuration of Abelian projected QCD.

From Fig.~\ref{fig:3dim}(a), 
one finds that the local chiral condensate tends to 
take a large value near the monopole 
location~\cite{Ohata:2020myj}.
Since monopoles appear on the dual lattice, 
we show the local monopole density $\rho_{\rm L}(s)$, 
as the average on closest dual sites. 
Of course, $\rho_{\rm L}(s)$ takes a large value near the monopole.
The distribution of the the local monopole density resembles that of the local chiral condensate,
as was pointed out in Ref.~\cite{Ohata:2020myj}.
Figures~\ref{fig:3dim}(c) and (d) show the Lorentz invariants 
${\cal F}$ and ${\cal G}$, respectively.
As a new result in this paper, 
we find that the distributions of 
${\cal F}$ and ${\cal G}$ also resemble that of the local chiral condensate.

The close relation of monopoles with 
${\cal F}$ and ${\cal G}$ might be understood, 
since the field strength tensor relates to monopoles
as $\partial_\mu \tilde F_{\mu\nu}^i=k_\nu^i$. 
Roughly speaking, the monopole can be a kind of source of 
${\cal F}$ and ${\cal G}$.
In contrast, their similarity with the local chiral condensate 
is fairly nontrivial.

In any case, 
we find clear correlation of distribution similarity
among the local chiral condensate, the local monopole density, 
and the Lorentz invariants ${\cal F}$ and ${\cal G}$
in Abelian projected QCD in the MA gauge.

\subsection{Correlation coefficients between local chiral condensate and magnetic variables}

In this subsection, we quantify the similarity 
between the local chiral condensate 
$\langle \bar \chi \chi (s) \rangle_u$ 
and magnetic variables, i.e., $\rho_{\rm L}(s)$, ${\cal F}(s)$ and ${\cal G}(s)$, defined in Section~\ref{Sect:magnetic}.
To this end, 
we use all the generated 100 gauge configurations 
in lattice QCD on $V = 24^4$ at $\beta = 6.0$, 
and calculate the local chiral condensate at
$2^4$ distant space-time points 
for each gauge configuration, 
resulting 1600 data points at each quark mass.

\begin{figure}[htb]
\centering
\includegraphics[width=14 cm]{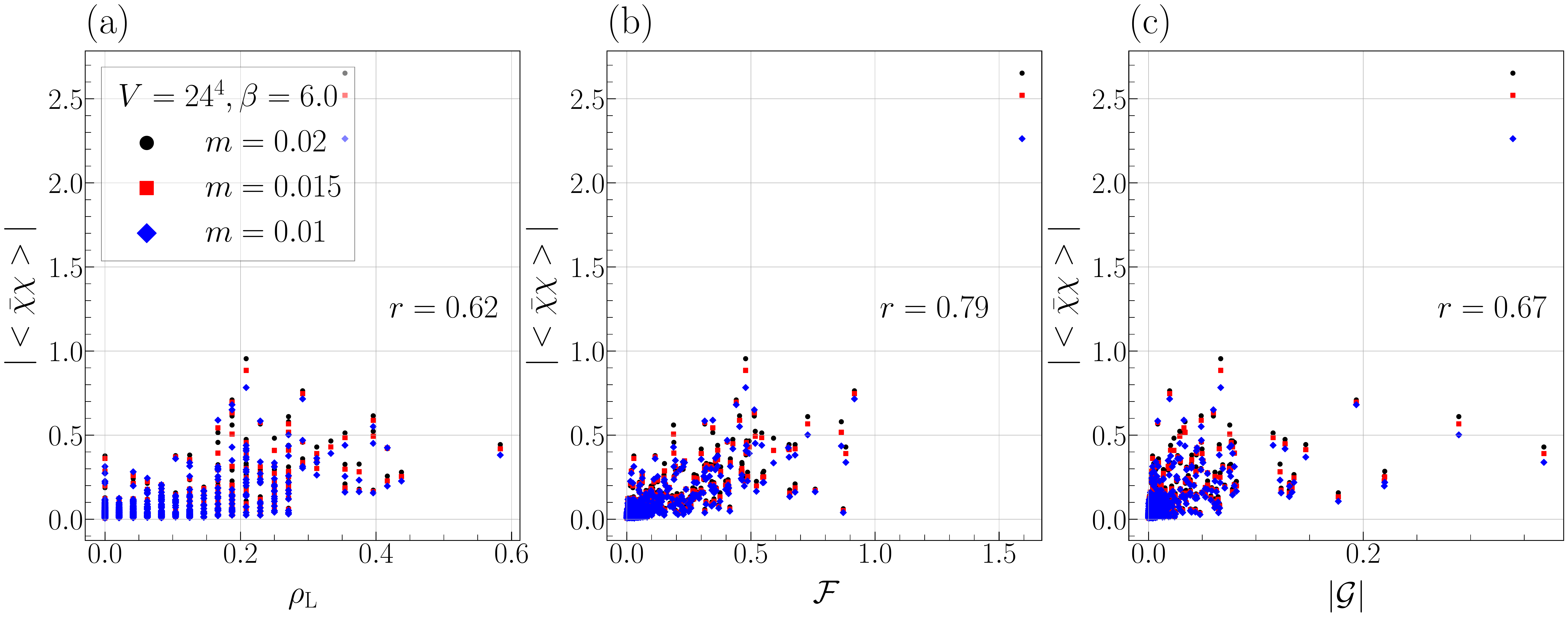}
\caption{The scatter plot between 
the local chiral condensate $\langle \bar \chi \chi (s) \rangle_u$ 
and (a) the local monopole density $\rho_{\rm L}(s)$, 
(b) ${\cal F}(s)$ and (c) $|{\cal G}(s)|$, 
using 100 Abelianized gauge configurations 
in SU(3) lattice QCD with $\beta=6.0$ and $V = 24^4$
at each quark mass $m$.} \label{fig:scatt}
\label{fig:scatter}
\end{figure}

Figure~\ref{fig:scatter} shows 
the scatter plot between 
the local chiral condensate 
$\langle \bar \chi \chi (s) \rangle_u$ 
and magnetic variables, i.e., 
the local monopole density $\rho_{\rm L}(s)$, 
Lorentz invariants ${\cal F}(s)$ and $|{\cal G}(s)|$, respectively, using 100 gauge configurations of 
Abelian projected QCD in the MA gauge, 
with the quark mass of $m=0.01, 0.015, 0.02$ 
in the lattice unit. 
In Fig.~\ref{fig:scatter}, positive correlation is qualitatively found 
between the local chiral condensate and the magnetic variables, $\rho_{\rm L}$, ${\cal F}$ and $|{\cal G}|$, respectively. 

Next, we consider a quantitative analysis
using correlation coefficients 
between the local chiral condensate and the magnetic variables, as a statistical indicator of correlation.
%, i.e., 
%$\rho_{\rm L}$, ${\cal F}$ and $|{\cal G}|$, %respectively.
%
In general, 
for arbitrary two statistical ensembles 
$\{A_i\}$ and $\{B_i\}$,
their correlation coefficient $r$ 
is defined as
\begin{eqnarray}
r \equiv \frac{\langle (A- \langle A \rangle)(B- \langle B\rangle)\rangle}{\sigma_A \sigma_B},
\end{eqnarray}
using the average notation $\langle ~~ \rangle$ 
and the standard deviation $\sigma_A \equiv \sqrt{\langle (A-\langle A \rangle)^2\rangle}$ and 
$\sigma_B \equiv \sqrt{\langle (B-\langle B \rangle)^2\rangle}$.
Here, $r=1$ means perfect positive linear correlation, and 
$r \gtrsim 0.7$ indicates strong positive linear correlation.

We measure correlation coefficients 
between the local chiral condensate 
$\left| \langle {\bar{\chi} \chi} (s) \rangle_u \right|$
and three magnetic variables, $\rho_{\mathrm{L}}(s)^{\alpha}$, $\mathcal{F}(s)^{\alpha}$ and $\left|\mathcal{G}(s)\right|^{\alpha}$, 
at various exponent $\alpha$, 
using 100 gauge configurations of 
Abelian projected QCD in SU(3) lattice QCD 
at $\beta=6.0$ on $V = 24^4$, 
for the quark mass $m =0.01$ in the lattice unit.
Table~\ref{table:CorrCoeffi} shows the result for the correlation coefficients.

\begin{table}[htb]
\centering
\begin{tabular}{ccccc}
\hline \hline
Lattice & $\alpha$ & $\rho_{\mathrm{L}}^{\alpha}$ & $\mathcal{F}^{\alpha}$ & $\left|\mathcal{G}\right|^{\alpha}$ \\ 
\hline
$V = 24^4, \, \beta = 6.0$
&$0.25$ & $0.47$ & $0.63$ & $0.60$ \\
& $0.5$ & $0.55$ & $0.71$ & $0.67$ \\
& $1$   & $0.62$ & $0.79$ & $0.67$ \\
& $1.5$ & $0.63$ & $0.81$ & $0.60$ \\
& $2$   & $0.60$ & $0.80$ & $0.55$ \\
\hline \hline
\end{tabular}
\caption{
Correlation coefficients between the local chiral condensate $\left| \langle {\bar{\chi} \chi} (s) \rangle_u \right|$
and three magnetic variables, $\rho_{\mathrm{L}}(s)^{\alpha}$, $\mathcal{F}(s)^{\alpha}$ and $\left|\mathcal{G}(s)\right|^{\alpha}$ 
at various $\alpha$, using 100 gauge configurations 
of Abelian projected QCD in SU(3) lattice QCD at 
$\beta=6.0$ on $V = 24^4$, 
for the quark mass $m =0.01$ 
in the lattice unit. 
}
\label{table:CorrCoeffi}
\end{table}

Quantitatively, the magnetic quantity ${\cal F}$ 
has the strongest correlation with the chiral condensate 
rather than $\rho_{\rm L}$ and ${\cal G}$.
As a conclusion of this paper, 
we find a strong positive correlation 
of $r\simeq 0.8$ between the local chiral condensate 
$\left| \langle {\bar{\chi} \chi} (s)\rangle_u \right|$
and the magnetic quantity ${\cal F}(s)$ 
in the confined vacuum of Abelian projected QCD.

\subsection{High-temperature deconfined phase}

Finally, we investigate also 
a high-temperature deconfined phase 
in lattice QCD on $V=24^3 \times 6$ at $\beta = 6.0$, 
where the temperature is $T\simeq$ 330 MeV above the critical temperature. 
We generate 200 gauge configurations, 
and calculate the local chiral condensate at
$2^3$ distant space points at a time slice for each gauge configuration,
resulting 1600 data points at each quark mass.

\begin{figure}[htb]
\centering
\includegraphics[width=14 cm]{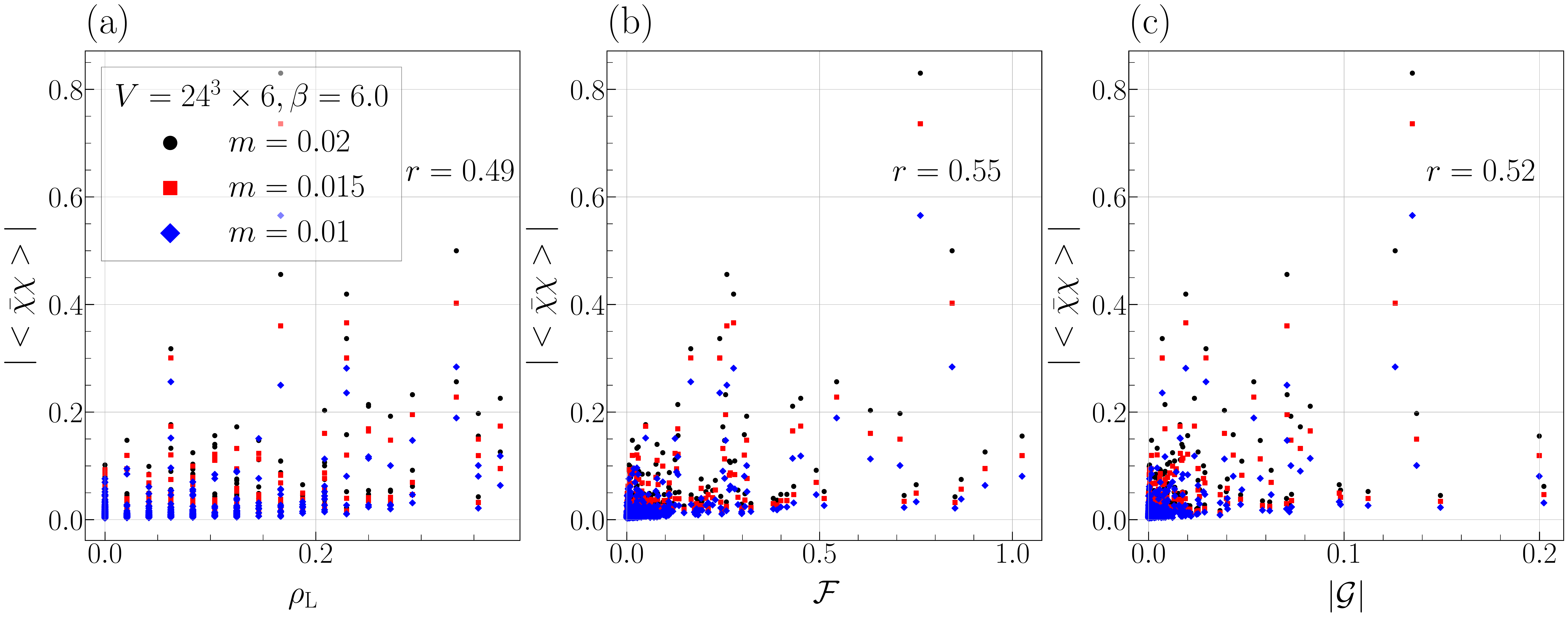}
\caption{
Result of the high-temperature deconfined phase 
for the scatter plot between 
the local chiral condensate $\langle \bar \chi \chi (s) \rangle_u$ 
and (a) $\rho_{\rm L}(s)$, 
(b) ${\cal F}(s)$ and (c) $|{\cal G}(s)|$, 
using 200 Abelianized gauge configurations 
in SU(3) lattice QCD with $\beta=6.0$ and $V = 24^3 \times 6$
at each quark mass $m$.}
\label{fig:scatt_high}
\end{figure}

Figure~\ref{fig:scatt_high} 
shows the scatter plot between 
the local chiral condensate 
$\left| \langle {\bar{\chi} \chi} (s) \rangle_u \right|$
and magnetic variables, i.e., 
the local monopole density $\rho_{\rm L}(s)$, 
and Lorentz invariants ${\cal F}(s)$ and $|{\cal G}(s)|$, respectively, using 200 gauge configurations of 
Abelian projected QCD in the MA gauge, 
with the quark mass of $m=0.01, 0.015, 0.02$ 
in the lattice unit. 

We show in Table~\ref{table:CorrCoeffi_high}
correlation coefficients 
between the local chiral condensate 
$\left| \langle {\bar{\chi} \chi} (s) \rangle_u \right|$
and three magnetic variables, $\rho_{\mathrm{L}}(s)^{\alpha}$, $\mathcal{F}(s)^{\alpha}$ 
and $\left|\mathcal{G}(s)\right|^{\alpha}$, 
at various exponent $\alpha$, using 200 gauge configurations 
of Abelian projected QCD in SU(3) lattice QCD at 
$\beta=6.0$ on $V = 24^3 \times 6$, 
for the quark mass $m =0.01$ in the lattice unit.

\begin{table}[htb]
\centering
\begin{tabular}{cccccc}
\hline \hline
Lattice & $\alpha$ & $\rho_{\mathrm{L}}^{\alpha}$ & $\mathcal{F}^{\alpha}$ & $\left|\mathcal{G}\right|^{\alpha}$ \\ 
\hline
$V = 24^3 \times 6, \, \beta = 6.0$
&$0.25$ & $0.37$ & $0.49$ & $0.49$ \\
&$0.5$  & $0.43$ & $0.55$ & $0.55$ \\
&$1$    & $0.49$ & $0.55$ & $0.52$ \\
&$1.5$  & $0.50$ & $0.51$ & $0.45$ \\
&$2$    & $0.49$ & $0.46$ & $0.38$ \\
\hline \hline
\end{tabular}
\caption{
Correlation coefficients in the deconfined 
phase between the local chiral condensate $\left| \langle {\bar{\chi} \chi} (s)\rangle_u \right|$ and three magnetic variables, $\rho_{\mathrm{L}}(s)^{\alpha}$, $\mathcal{F}(s)^{\alpha}$ and $\left|\mathcal{G}(s)\right|^{\alpha}$ 
at various $\alpha$ in Abelian projected QCD of 
SU(3) lattice QCD 
at $\beta=6.0$ on $V = 24^3 \times 6$ for $m=0.01$.}
\label{table:CorrCoeffi_high}
\end{table}

From Fig.~\ref{fig:scatt_high} and Table~
\ref{table:CorrCoeffi_high}, 
all the correlations between the local chiral condensate 
and the three magnetic variables, 
$\rho_{\rm L}$, ${\cal F}$ and ${\cal G}$, 
become weaker in the deconfined phase, 
where the chiral condensate itself goes to zero in the chiral limit. 

\section{Summary and Conclusion}

We have studied the relation among the local chiral condensate, monopoles, and magnetic fields, using the lattice gauge theory, as a continuation of Ref.\cite{Ohata:2020myj}.

First, we have created idealized Abelian gauge systems of 
1) a static monopole-antimonopole pair and 
2) a magnetic flux without monopoles, 
on a four-dimensional Euclidean lattice.
In these systems, we have calculated the local chiral condensate on quasi-massless fermions 
coupled to the Abelian gauge field, and have found that the chiral condensate is localized 
in the vicinity of the magnetic field. 

Second, performing SU(3) lattice QCD Monte Carlo simulations, 
we have investigated Abelian projected QCD 
in the maximally Abelian gauge, 
and have found clear correlation of distribution similarity 
among the local chiral condensate, color monopoles, and color magnetic fields in the Abelianized gauge configuration.

As a statistical indicator, 
we have measured the correlation coefficient $r$, 
and have found a strong positive correlation 
of $r \simeq 0.8$ between 
the local chiral condensate and the Euclidean 
color-magnetic quantity ${\cal F}$.

We have also examined the local correlation 
in the deconfined phase of thermal QCD, 
and have found that the correlation between 
the local chiral condensate and magnetic variables 
becomes weaker. 

Thus, in this paper, we have observed a strong correlation 
between the local chiral condensate and 
magnetic fields in both idealized Abelian gauge systems 
and Abelian projected QCD. 
From these results, we conjecture that 
the chiral condensate is locally enhanced by the strong color-magnetic field 
around the monopoles in Abelian projected QCD, 
like magnetic catalysis.

Note however that this correlation does not necessarily 
mean that chiral symmetry breaking 
is caused by the non-uniform magnetic field. 
To realize spontaneous chiral-symmetry breaking, 
as was discussed in subsection~\ref{subsec:analytic}, we need some zero mode in the denominator of $I$ in the chiral limit.
In the context of the dual superconductor picture, 
this might be realized by condensation of monopoles, 
as was suggested in the dual Ginzburg-Landau theory 
\cite{Suganuma:1993ps}.

To conclude, once chiral symmetry is spontaneously broken, 
the local chiral condensate is expected to have a strong 
correlation with the color magnetic field.

\begin{acknowledgments}
H.S. is supported in part by the Grants-in-Aid for
Scientific Research [19K03869] from Japan Society for the Promotion of Science. H.O. is supported by a Grant-in-Aid 
for JSPS Fellows (Grant No.21J20089).
Most of numerical calculations have been performed on OCTOPUS, at the Cybermedia Center, Osaka University. 
We have used PETSc to solve linear equations 
for the Dirac operator~\cite{petsc-web-page,petsc-user-ref,petsc-efficient}. 
\end{acknowledgments}
\bibliography{mag_cat}
\end{document}